# Desynchronization in multilayer *p-i-n* drift step recovery diode

Jorge Pena Lozano, Kindred Griffis, Dimash Aimurzayev, Gregory Wierzba, Sergey V. Baryshev

Department of Electrical and Computer Engineering, Michigan State University, East Lansing, MI, United States

*Abstract—* The impact semiconductor drift step recovery diodes (DSRDs) can have on high-power microwave applications make them a device of interest for future solid-state electronics. However, there is little known about their functionality and degradation under over-voltaging or high average power dissipation conditions that could therefore hinder their continual design and optimization toward better performance. The experiments on a Si seven layer DSRD conducted in the present paper allowed to broaden the understanding of its opening switching performance under over-voltaging conditions. A striking desynchronization was discovered and linked to junction and electro-neutral region damage through experiments and PSpice modelling.

## I. INTRODUCTION

The relentless pursuit of high-speed and high-power electronic systems has propelled advancements in wide bandgap (WBG) semiconductors, beyond traditional Si electronics [1], because WBG has superior operating power and frequency figures of merit. When measured and compared on the scale of so-called Johnson figure of merit (JFOM), indirect (and wide) bandgap materials like SiC and diamond hold the promise of outperforming Si by 10-100 times. At the same time, there exists a class of power avalanche *p-i-n* diodes (called diode avalanche shaper termed DAS), developed and matured particularly for Si, that is ~10× faster than JFOM [2]. Owing to excellent quality of Si and its electronic properties and overall device fabrication repeatability and yield, avalanche ultrafast Si-based *p-i-n* diodes remain a desired technology for fully solid-state short-pulse (≤100 ps) GW-class peak power drivers for advanced radars, combustion/chemical/biomedical systems and any advanced sub-100-ns pulse forming systems to drive or detect ultrafast processes [2-4].

In inductive storage pulse-forming circuits, an intermediate step between high power FET, producing initial high voltage nanosecond pulse, and the final avalanche shaper an intermediate device, called drift step recovery diode (DSRD) is required to properly trigger the avalanche streamer in DAS. The mating condition between DSRD and DAS is called the voltage ramp rate *dV/dt* that is exactly the JFOM [5]. The pulse forming generators use the DSRD and DAS that work in pair as opening and closing switches in order to modify and compress a pulse to several or tens of kilovolts and rise times near or shorter than 100 ps range.

Typically, for a higher voltage rating many *p-i-n* diodes are stacked together to form a DSRD. When such a stacked DSRD is switched on/off, it behaves as a whole, meaning there is synchronicity between all of the individual *p-i-n* diodes in the stack. In this work, we describe and discuss an effect of switching de-synchronization caused by a high voltage cycling of a 7-stack Si DSRD. This effect was modelled in PSpice and individual *p-i-n* diode parameters were extracted allowing for quantitative insights into this DSRD degradation mechanism.

## II. SAMPLES

The DSRD devices used in this study were *p+-i-n+* structures with intrinsic layer that doped with accepters at the level of $(2\text{-}3)\times10^{14}$ 1/cm$^3$. The high conductivity regions were doped through a high-temperature diffusion to create doping profiles optimized for high power switching. The resulting *p+-i-n+* structures were then metal coated for high quality ohmic contacts and stacked together via an eutectic bond. The resulting stack was laser cut as a circular DSRD device followed by hydrofluoric acid, nitric acid, and acetic acid etch to remove some of the silicon damage, and finally coated with polyimide for edge protection from breakdown. This production method and its variations and comparisons were published elsewhere [6] in greater detail. The tested DSRD and its custom compression holder are shown in Fig.1.

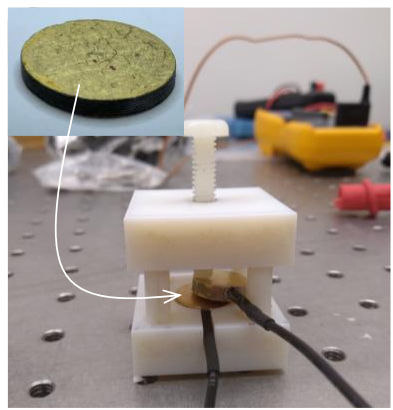

**Fig.1.** An image of a custom compression DSRD holder with the DSRD sample tested shown in the inset.

## III. METHODS

The diodes were measured in DC and AC to obtain *I-V* curves, ideality, saturation current, series resistance, junction capacitance, and reverse recovery time. Reverse *I-V* measurements and conditioning of the diodes are discussed first, since understanding conditioning is a prerequisite to understanding all other measurements.

### A. Diode current-voltage measurement

The standard process of curve tracing was used to measure the current-voltage (*I-V*) characteristics of the diode, both in forward and reverse directions. Forward curve was measured using a Keithley 2410 that could accurately measure the current in the nA range. Because the DSRDs were rated at a breakdown voltage of 3 kV and the Keithley 2410 was limited up to -1.1 kV, a separate Matsusada AU-5R12-LC limited up to -5 kV (12 mA and 60 W) was used for curve tracing in reverse, with current sensitivity in the μA range.

After obtaining the *I-V* curves, a pair of points $(V_1, I_1)$ and $(V_2, I_2)$ were used to extract the ideality factor $\eta$ and saturation current $I_s$ of the diode as

$$\eta = \frac{V_1 - V_2}{V_t \ln(I_2/I_1)} \quad (1)$$

$$I_s = I_1 e^{-V/\eta V_t} \quad (2)$$

where $V_t$ is the thermal voltage. The series resistance $R_s$ was also determined (where possible) as the ratio of voltage to current at the highest registered current, in the linear portion of the forward *I-V* curve, where a deviation from the ideal Shockley-type behavior was detected.

### B. Diode Junction Capacitance Measurement

In reverse-bias conditions, one important parameter when switching a diode is the depletion capacitance $C_j$, also known as junction capacitance. A Teledyne T3AFG40 function generator and the Tektronix MSO 2024B oscilloscope were used to measure $C_j$. A separate circuit (shown in Fig.2) was built to measure the complex impedance of the diode $Z_D = R + jX$. The function generator is connected between the diode cathode and a 50 Ω resistor. AC signal of 200 mV$_{pp}$ at 300 kHz and DC offset between 0 and -10 V is applied. Because in the reverse, the diode act as a variable capacitor (when different DC biases are applied), input and output sine signals measured as Ch1 and Ch2 provide the phase shift which allows for calculating the impedance and $C_j$.

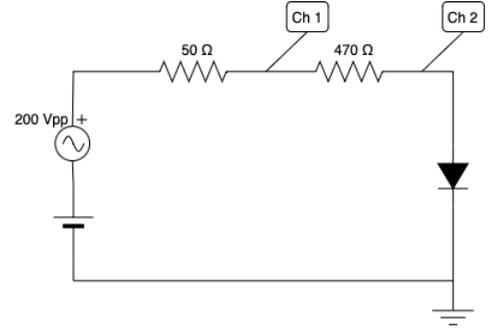

**Fig.2.** Circuit for junction capacitance measurements.

The calculation procedure was as follows. After having the voltage and phase measured at Ch1 and Ch2 at DC bias zero, their relationship is

$$V_1 \angle \theta_1 = V_2 \angle \theta_2 \cdot \frac{Z_D}{Z_D + 470\Omega} \quad (3)$$

where $V_1 \angle \theta_1$ and $V_2 \angle \theta_2$ are the voltage and phase around the 470 Ω resistor, and $Z_D$ is the diode impedance. After finding this value, the probe capacitance (2.2 pF) was also accounted for to find $C_{j0}$ and the rest of capacitances. From that, voltage dependent ($V_D$) capacitance $C_j$ was modeled as

$$C_j = \frac{C_{j0}}{(1 - V_D/\Phi_0)^m} \quad (4)$$

where $C_{j0}$ is the zero-bias junction capacitance, $\Phi_0$ is the calculated built-in voltage and $m$ is the grading coefficient. From the measured data $C_j$ is deduced and the rest of the parameters can be extracted as

$$\Phi_0 = \frac{V_{D1}V_{D2} - V_{D3}^2}{V_{D1} + V_{D2} - 2V_{D3}} \quad (5)$$

$$m = \frac{\ln(C_{j1}/C_{j2})}{\ln((\Phi - V_{D2})/(\Phi - V_{D1}))} \quad (6)$$

where $C_{j1}$, $C_{j2}$, and $C_{j3}$ are measured junction capacitance at $V_{D1}, V_{D2}, V_{D3}$ reverse voltages respectively such that that $C_{j3} = \sqrt{C_{j1}C_{j2}}$.

### C. Reverse Recovery Measurements

The reverse recovery transients (on-to-off time $\tau_s$) were measured by applying a forward pulse followed by an abrupt change in polarity to a reverse pulse. The forward bias pushes minority carriers across the junction as forward current flows. After switching to reverse bias, a reverse current flows until the diode has emptied from the charge and then it turns off. The measurements were taken with an Avtech AVR-EB4-B pulser such that the forward-to-reverse current ratio ($I_f/I_r$) was changed from 2 to 1/12. The power diode was placed in a custom holder. Waveforms were captured using a LeCroy SDA6000 oscilloscope. Finally, $t_s$ was converted to the minority carrier lifetime $\tau_p$ as [7]

$$t_s = \tau_p \cdot ln\left(1 + \frac{I_f}{I_r}\right) \quad (7)$$

## IV. RESULTS

### A. Reverse I-V

The DSRD reverse current was originally measured at voltages up to -1.1 kV, during which the reverse current did not exceed 300 µA (stage 1 in Fig.3). The measurement was repeated with the intent to gradually condition the diode to higher voltages until breakdown voltage mark was approached. During stage 2, up to -1.7 kV was applied. The onset of a pre-breakdown was observed with current increased to about 2.4 mA. This sweep was terminated in order to consider the finding before repeating the experiment or apply higher voltages, as this voltage (giving rise to mA leakage current) was much smaller than the expected breakdown voltage. In the next voltage sweep, the diode unexpectedly exhibited currents in the low mA range at just above 500 V (stage 3). Every consecutive sweep led to progressively higher leakage current and lower voltages (stages 4 and 5).

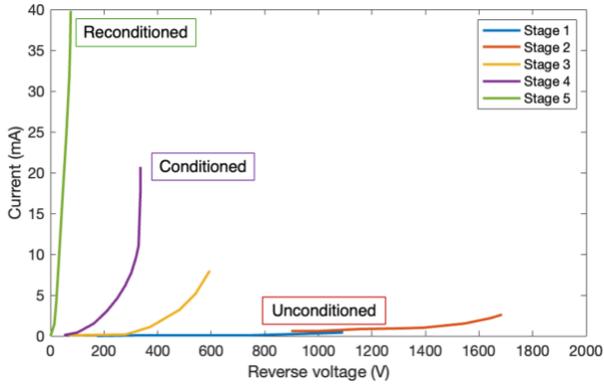

**Fig.3.** Reverse curve tracing and corresponding conditioning of the diode.

At stage 5, the DSRD was found to have a breakdown voltage of 20 V, at which point reverse current was ~5 mA. In what follows, we use terms *unconditioned* (all DC, AC or switching tests carried out before stage 1), *conditioned* (all DC, AC or switching tests carried out after stage 3), *reconditioned* (all DC, AC or switching tests carried out after stage 5).

### B. Forward I-V

The forward bias curve tracing measurements are shown in Fig.4, from were $\eta$, $I_s$ and $R_s$ were extracted as summarized further in Table I. It can be clearly seen that (reverse bias) conditioning changed the forward bias characteristics in that after conditioning it takes more voltage to attain the same current. In other words, $\eta$ increases at a simultaneous drastic increase in saturation current $I_s$. For the unconditioned diode, the series resistance could not be calculated because the equipment used could not supply high enough current to detect a voltage drop indicating the transition from diode to resistor-like behavior. This suggests that unconditioned $R_s$ was much lower than 1 Ω.

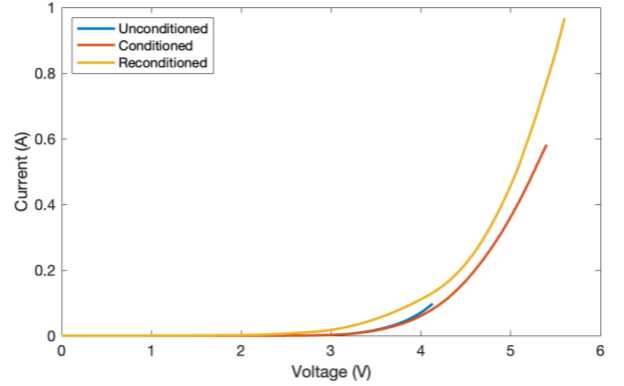

**Fig.4.** Forward *I-V* curves of the DSRD, measured.

TABLE I

PARAMETERS FOR ALL CASES

| Diode | $\eta$ | $I_S$ | $R_S$ |
|---|---|---|---|
| Unconditioned | 9.205 | 7.3 nA | - |
| Conditioned | 16.96 | 6.6 µA | 1.825 Ω |
| Reconditioned | 21.154 | 74. 1µA | 1.418 Ω |

### C. Junction Capacitance

*C-V* curves shown in Fig.5 underscore a strong effect of conditioning on the junction capacitance of the DSRD. Here experimental points are fit by least-square method using Eq.(4) allowing for determining built-in potential and the grading coefficient, as summarized in Table II. It can be seen that after conditioning the junction capacitance has decreased slightly and then, after reconditioning, greatly increased in magnitude, suggesting a significant change of the junctions of the DSRD.

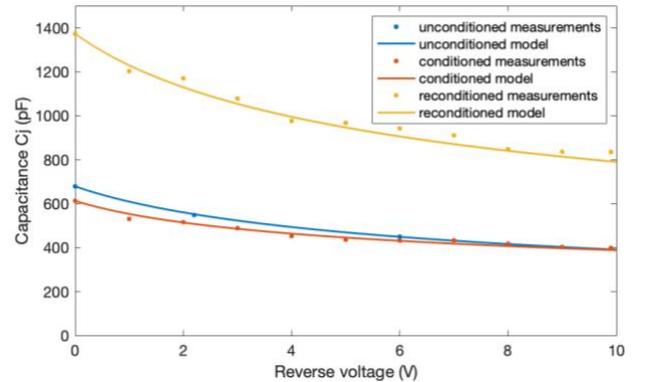

**Fig.5.** Junction capacitance of the DSRD, measured and modeled.

## TABLE II
CAPACITANCE MODEL PARAMETERS FOR ALL CASES

| Diode | $C_{j0}$ | $\Phi$ | $m$ |
|---|---|---|---|
| Unconditioned | 678.7pF | 3.025 V | 0.378 |
| Conditioned | 613.0pF | 2.022 V | 0.256 |
| Reconditioned | 1372.03pF | 2.833V | 0.3658 |

### D. Reverse recovery

ON/OFF diode switching was measured at varied forward-to-reverse current ratio, from 2/1 to 1/12, to visualize the waveforms and to calculate $\tau_p$. The summary of unconditioned, conditioned, and reconditioned states of the DSRD are shown in Fig.6. The curves were normalized in order to show the details and differences between performance under various current ratios. Note, the ringing in the measurement of the unconditioned diode was caused by the inductance of initially too long 18 AWG wires connecting the diode holder and the Avtech pulser. Later, when measuring the conditioned diode, wires were made shorter and ringing quenched. For unconditioned and conditioned there little to none difference in terms of switching behavior.

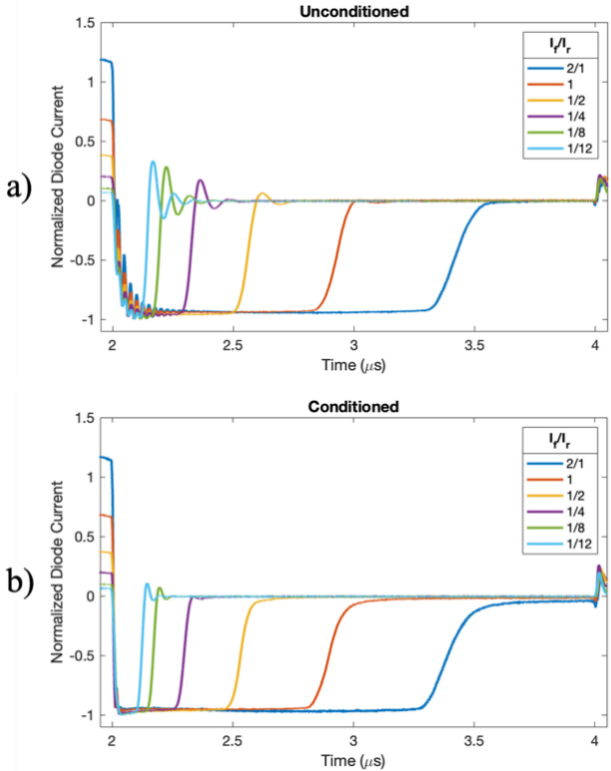

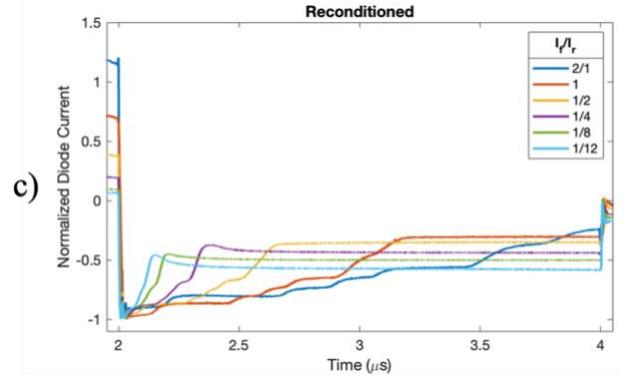

**Fig.6.** Switching waveforms of the DSRD through conditioning at varying ratios $I_f/I_r$.

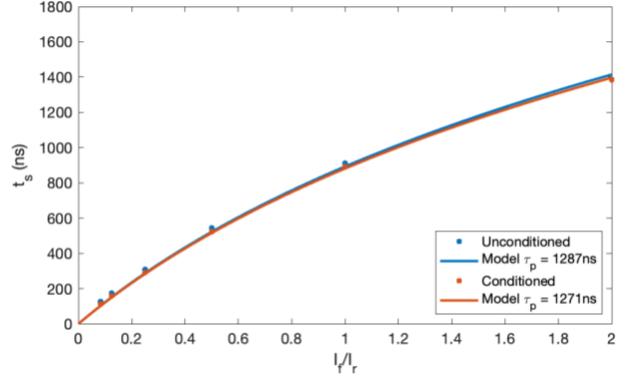

**Fig.7.** Switching time as function of the switching ratio to calculate $t_p$.

From the measured recovery time $t_s$ within the used forward-to-reverse range, Eq.(7) yields the minority carrier lifetime $\tau_p$ of 1.287 μs for unconditioned case which is a robust number for a power diode switch [8]. Carrier lifetime slightly increased (within the margin of error) after conditioning, as seen in Fig.7. Within the tested reversed voltage range (Fig.6b), unconditioned-to-conditioned the switching behavior did not change much despite the visible change in DC properties (Table I) and capacitance parameters (Table II). After reconditioning however, significant modifications to DSRD took place (captured in Table I and II and Fig.4 and 5). As seen in Fig.6c, most striking in switching are 1) the device began to behave like a resistor after recovery, whereby it did not closed off after emptying of charge but kept conducting in reverse until the negative voltage pulse ended; and 2) the stepped-like behavior emerged, whereby while recovering seven different steps could be seen in the measurements (more prominent for larger ratios) pointing towards a desynchronization between the individual diodes in the stacked device.

## V. DISCUSSION

Based on the described experimental data, a PSpice model was developed to model the switching behavior as *MODEL DSRD D (Is + N + Rs + Cj0 + M + Vj + BV + IBV + Tt)* where parameters respectively correspond to saturation current, ideality factor, zero-bias junction capacitance, grading coefficient, built-in potential, breakdown voltage, breakdown current, and $Tt \equiv \tau_p$ when synchronized and $Tt$ is the cumulative $\tau_p$ when out of sync. This model was then adapted to represent the DSRD as seven individual diodes (Fig.8) to pinpoint the degradation mechanism.

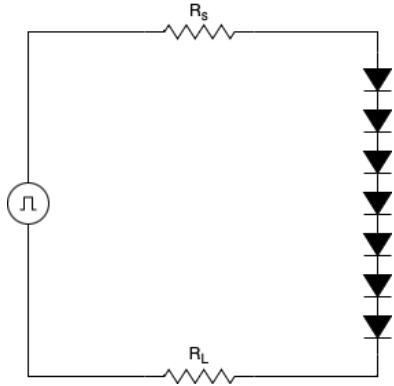

**Fig.8.** Circuit for PSpice model of 7-layered DSRD.

From waveforms shown in Fig.6c, it can be seen that the diodes are not working in sync anymore, and the difference in the basic parameters were hypothesized to cause the observed behavior. Fig.9 presents a schematic showing how the basic *MODEL DSRD* parameters affect diode switching: $Tt$ sets the bottom flat of the pulse, transitions between the steps are charted by $Cj0$, the step heights are defined by the breakdown voltage of each *p-i-n*.

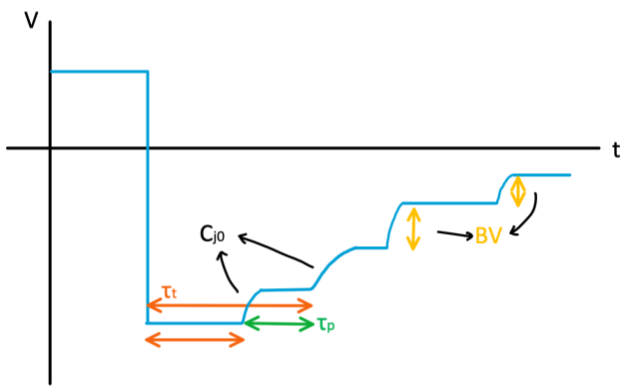

**Fig.9.** *.MODEL DSRD* parameter effects on reverse recovery transient.

The first step to test established *.MODEL DSRD* was to match it with the measured data in PSpice for the unconditioned state of the DSRD. In this case, all the diodes were to have the same parameters as listed in Tables I and II. In Fig.10, the measured data in blue is compared to the model response in orange at varied $I_f/I_r$ ratios of 2:1, 1:1, 1:2, 1:4, 1:8, 1:12. It can be seen, the higher the ratio, the more discrepancy there is. The higher the forward voltage compared to the reverse one, the least accurate the model is. The reason behind it is that the model does not vary $Tt$ which is locked at 1,287 ns as obtained from the fitting in Fig.7, but one can see that there is a variation in the data points. All other parameters capture the switching behavior perfectly especially as seen for the small ratio data.

Next, the model was reconfigured and split into seven individual diodes (Fig.8) to capture the desynchronization. Modelling of the reconditioned DSRD was carried out for three ratios: 1/2, 1/1, and 2/1. This is because at smaller ratios bottom flat sections were blurred by transitions sections due to strongly enhanced capacitance $Cj0$ (as shown in Fig.5). $R_s$ was used as measured and reported in Table I and kept the same. For modeling unconditioned, conditioned and reconditioned states of DSRD, the built-in potential and grading coefficient were updated according to the values reported in Table II in all three models. All individual diode's properties were fine-tuned (in accord with the concept presented in Fig.9) to recover the experimental waveforms as closely as possible. It was found that the breakdown voltage ($BV$) and transit lifetimes ($\tau_p$) were different for each layer.

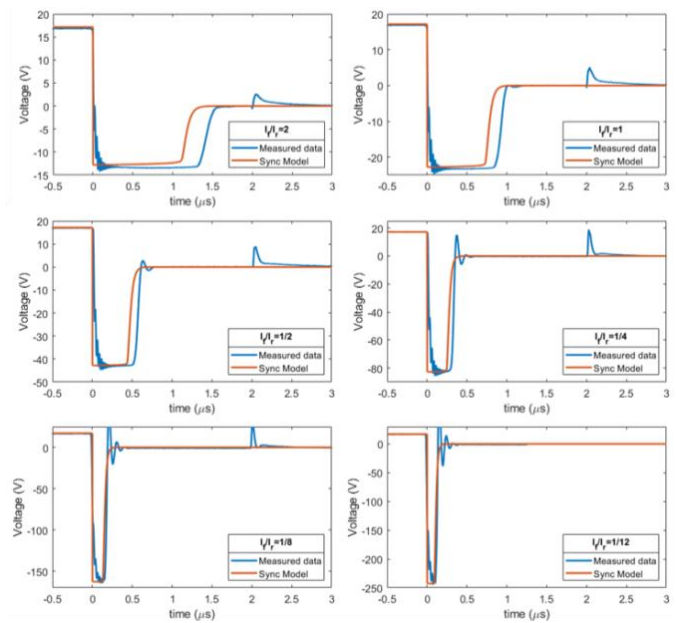

**Fig.10.** PSpice model results vs measured data at different forward to reverse ratios (as labeled).

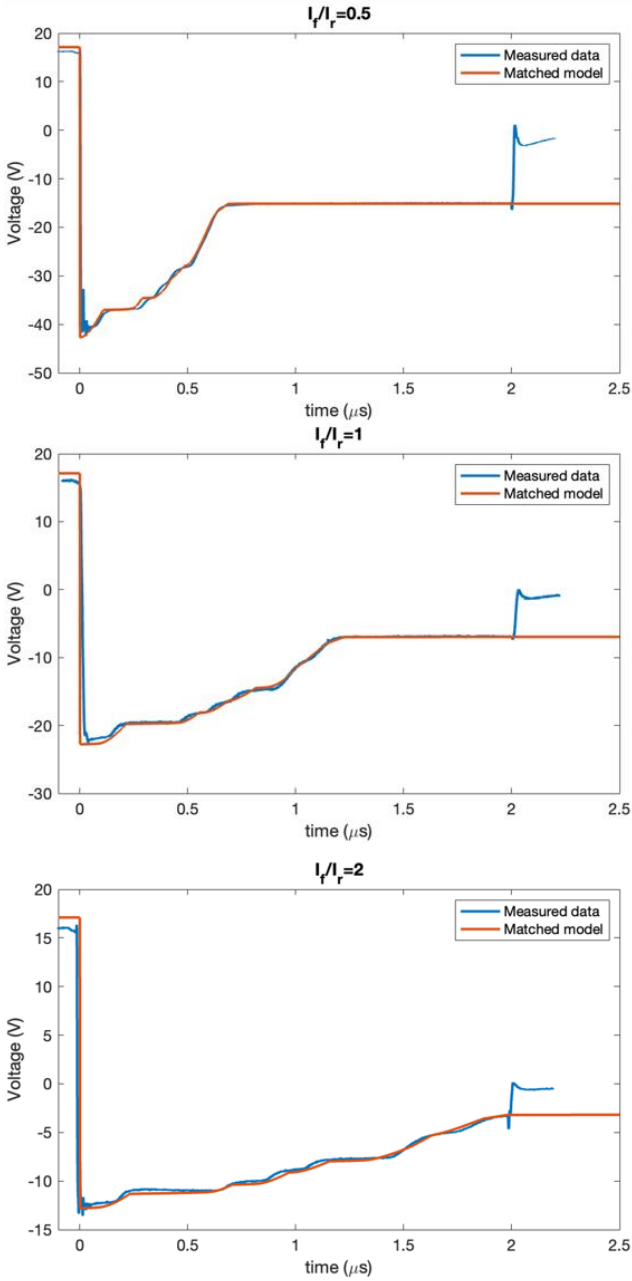

**Fig.11.** Closest match models for each switching ratio and the measured data.

TABLE III

BREAKDOWN VOLTAGE FOR EACH DIODE LAYER, PER MODEL

| Ratio | Breakdown Voltage (V) | | | | | | | |
|---|---|---|---|---|---|---|---|---|
|  | D1 | D2 | D3 | D4 | D5 | D6 | D7 | Total |
| 0.5 | 52 | 19 | 27 | 29 | 45 | 50 | 25 | 427 |
| 1 | 25 | 11 | 13 | 15 | 33 | 19 | 13 | 227 |
| 2 | 10 | 4 | 8 | 8 | 21 | 10 | 4 | 128 |
| DC |  |  |  |  |  |  |  | 20 |

TABLE IV

MINORITY CARRIER LIFETIME FOR EACH DIODE LAYER, PER MODEL

| Ratio | $\tau_p$ (µs) | | | | | | |
|---|---|---|---|---|---|---|---|
|  | D1 | D2 | D3 | D4 | D5 | D6 | D7 |
| 0.5 | 0.1 | 0.55 | 0.23 | 0.12 | 0.2 | 0.1 | 0.05 |
| 1 | 0.2 | 0.55 | 0.18 | 0.15 | 0.27 | 0.15 | 0/unresolved |
| 2 | 0.2 | 0.45 | 0.23 | 0.17 | 0.35 | 0.15 | 0.05 |

Tables III and IV summarize the specific values for each $\tau_p$ and $BV$, differential for the seven different p-i-n layers for three switching ratios. In Table III, the voltages at which each device begin leaking is highlighted. The total $BV$ on the last column is highest reverse voltage, as seen in Fig.11. The smaller the switching ratio, the faster the extraction of charge, and the higher the apparent breakdown voltage, changing from 427 V down to 128 V, down to approx. 20 V under DC conditions (Fig.3). When such a dependence exists, it suggests internal junction damage. In such scenario, a non-avalanche breakdown process takes place that is a result of physical defects affecting the efficiency of carrier removal [9]. Results in Fig.3 help connect the dots and explain the observed loss of performance. After measuring the forward I-V characteristics of the unconditioned, conditioned, and reconditioned DSRD, an increase in $\eta$ and $I_s$ were seen from state to state which indicates degradation of the physical properties of the device.

Together, it appears as if conditioning created carrier traps within the device's electroneutral regions which would serve as centers/sinks for carrier recombination and/or increased conduction by defects allowing leakage current paths, thus increasing both the saturation current and the ideality factor [10, 11]. This explains very well, in some layers, a more than a ten-fold decrease in the minority carrier lifetimes (Table IV). Indeed, the carrier lifetime results extracted from fitting the experimental waveforms point toward the same conclusion. The highest values of about 500 ns exist for layer D2, while lowest are from 50 ns to unresolvable were found for layer D7.

Another co-existing possibility, suggested by the capacitance measurements, could be degrading junctions, where a reduced barrier potential would also explain higher currents at lower voltages [12]. The device also became more capacitive as seen in Fig.5. From the unconditioned to the reconditioned measurements, the junction capacitance doubled. This increase can be explained by a narrowing depletion region width, which matches with the effect of lower built-in potential in Table II and further points to a catastrophic junction damage explained above. Given the stacking nature of the 7-layer DSRD, additionally extrinsic damages such as in the metal contacts of the diode could have taken place that would cause $R_S$ to increase. However, this is expected to play a minimal role in the present case.

## VI. CONCLUSION

In conclusion, the experiments on a multilayer DSRD conducted in the present paper allowed to broaden the understanding of its performance through analyzing degradation regimes. When conditioned using reverse voltage sweeps, the primary features discovered are

1) Remarkably lowered breakdown voltage;

2) Weakened/failed performance to act as the fast-opening switch as it requires high reverse voltages to remove carriers from the junctions;

3) Striking desynchronization between all 7 *p-i-n-* layers.

The potential reasons of degradation were categorized as intrinsic and extrinsic to the diode. The former implied the formation of defects within the electroneutral regions and *p-i-n* junctions where the defect effects are signified by increased leakage current, ideality factor and capacitance, and reduced built-in potential. Using PSpice, degraded performance was analyzed and quantified. These results further corroborated the importance of intrinsic effects by revealing non-avalanche breakdown behavior and demonstrating dire decrease of carrier lifetime, from 1.2 μs to less than 0.5 μs; down to 50 ns for some of the *p-i-n* layers comprising the DSRD. On the other hand, extrinsic effects would refer to the diode contact leads and metals within each layer, and its degradation could cause increased series resistance.